\documentclass[aps,amssymb,showpacs,floatfix, bibliography, twocolumn]{revtex4-1}
\usepackage{amssymb,graphicx}
\usepackage{epsfig,graphicx}
\usepackage{subcaption}
\usepackage{subfloat}
\usepackage{epstopdf}
\usepackage{amsmath}
\usepackage{tikz}

\begin{document}

\title{Hopfions in lattice dimer model}

\author{Grigory Bednik}
\affiliation{ UC Santa Cruz Physics Department, 1156 High Street
Santa Cruz, CA 95064, USA}


\newcounter{TypeOne}
\newcounter{TypeTwo}
\newcounter{TypeThree}

\setcounter{TypeOne}{1}
\setcounter{TypeTwo}{2}
\setcounter{TypeThree}{3}

\begin{center}
\begin{abstract}

In this paper, we study topological properties of 3D lattice dimer model. We demonstrate, that the dimer model on a  bipartite lattice  possesses topological defects, which are \textit{exactly} characterized by Hopf invariant. We derive its explicit algebraic expression in terms of effective magnetic field of a dimer configuration. Thus, we solve the problem of topological classification of possible states in 3D lattice dimer model. Furthermore, since the lattice dimer model is known to be dual to spin ice, our work can be viewed as a proposal to search for hopfions in classical, as well as, artificial spin ice and related materials.

\end{abstract}
\end{center}

\maketitle

\section{Introduction}

Lattice dimer model has a very long history in condensed matter physics. Its idea is that lattice can be viewed as a graph, such that its edges connecting nearest neighboring sites, are either empty or filled with dimers, so that every vertex is connected to exactly one dimer.  Lattice dimer model was first proposed back in 1960-s as a tool to solve 2D Ising model \cite{doi:10.1063/1.1703953} - it was found that partition function of the latter is related to the number of dimer coverings in a dual dimer model, which on a planar lattice can be computed exactly. Lattice dimer model was extensively used in an attempt to describe high-temperature superconductivity: it was used to represent resonant valence bond state, where electrons occupying neighboring sites form pairwise singlets \cite{ANDERSON1196}. In this context, lattice dimer model was generalized to quantum dimer model \cite{PhysRevLett.61.2376}, where quantum mechanical state can be viewed as a superposition of dimer coverings, and quantum evolution consists of local dimer flips. It was found, that such model possesses several non-trivial phases: in different regimes, its ground state can form either classical trivial, or staggered phase, or quantum RVB phase, which may be gapless $U(1)$ phase on a bipartite lattice, or a gapped $\mathbb{Z}_2$ phase, if the underlying lattice is non-bipartite. 

The fact that quantum dimer model on a bipartite lattice hosts $U(1)$ phase is not accidental. As was shown in Ref. \cite{PhysRevLett.91.167004}, dimer configurations on a bipartite lattice can be described using effective magnetic field, which, in turn, can be represented in terms of effective vector potential, similarly to conventional $U(1)$ gauge field. In two-dimensional case, such vector potential is reduced to, so called, height representation (see \cite{lacroix2011introduction} for review), but in three dimensional case, the vector potential is an actual vector, as in the case of physical electromagnetic field.

Since classical lattice dimer model can be viewed as an appropriate limit of quantum model, its evolution occurs due to local flips of dimers along a plaquette. Thus, a natural question to ask is, whether classical lattice dimer model hosts different topological sectors, which cannot be connected to each other through local flips. This question was considered, e.g. in Ref.  \cite{PhysRevB.84.115129}, where, using exact diagonalization, it was shown that dimer model on a diamond lattice contains different topological sectors. 
 In addition, possible topological sectors in 3D dimer model on a cubic lattice were also explored in \cite{PhysRevB.84.245119}: it was shown that, in the continuum limit, dimer model behaves as $O(n)$ field, which has topologically non-trivial configurations called \textit{hopfions}. Furthermore, in the Ref \cite{PhysRevB.84.245119}, there was presented an explicit dimer configuration, which, in the continuum limit, becomes a hopfion.

In continuum field theory, hopfions exist, because three-dimensional vector field may have different topological sectors characterized by Hopf invariant (see \cite{arnold2013topological}, and also \cite{e1996force, barrett1995advanced}). Such invariant can, for example, describe skyrmions in $O(3)$ vector model \cite{PhysRevLett.51.2250}. Hopf invaiant was first proposed to describe topological defects in liquid helium \cite{Volovik1977}, but it attracted a lot of attention more recently,
 e.g. it was found that Hopf invariant can be responsible for constraint on plasma relaxation \cite{1742-6596-544-1-012006}, and, more interestingly, it can lead to new kinds of topological phases of matter \cite{PhysRevB.88.201105, PhysRevB.95.161116, PhysRevB.94.035137, PhysRevB.96.041103}.

In this paper, we demonstrate presence of topological defects with non-trivial Hopf number in a dimer model on a cubic lattice. Specifically, we consider topologically non-trivial dimer configurations, presented in \cite{PhysRevB.84.245119}, which were proven to become hopfions in continuum limit, and claim, that they carry an exact Hopf number on a lattice. The latter can be evaluated by considering the effective magnetic field of the corresponding lattice, and computing a discretized expression for its Chern-Simons integral similarly to the case of skyrmions  \cite{PhysRevLett.51.2250}.

This paper is organized as follows. In Sec. \ref{EffectiveMagneticField}, we revisit the concept of effective magnetic field on a dimer lattice and introduce its new definition 
 suitable for describing topological defects. In Sec. \ref{EffectiveVectorPotentialAndLatticeHopfNumber}
we introduce our Hopf invariant and describe its main properties. In Sec. \ref{Discussion} we summarize our results and discuss their possible applications. A few technical details are discussed in the appendix.


\section{Effective magnetic field}
\label{EffectiveMagneticField}

We start from revision of basic properties of lattice dimer model. We consider 3D cubic lattice, which nearest neighboring vortices are connected with an edge. Every edge can be either empty or occupied with a dimer in such a way, that every vertex is connected to exactly one dimer. Evolution of a dimer configurations can be realized through local flips: if a plaquette has two aligned dimers, they can simultaneously change their directions: 


\begin{figure}[h]
\includegraphics{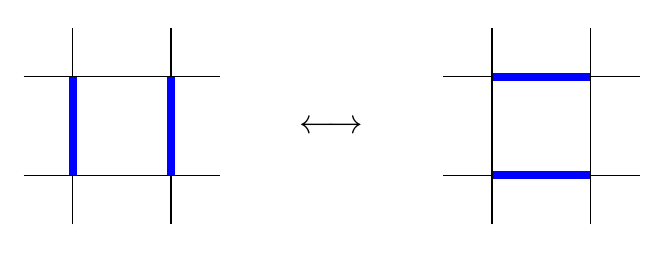}
\caption{A local dimer flip in lattice dimer model.
}
\label{DimerFlip}
\end{figure}

Let us revisit the construction of effective magnetic field on a general bipartite dimer lattice. As we mentioned previously, we have a constraint, that 
every vertex is attached to exactly one dimer. In other words, we say that sum of occupation numbers over all edges attached to a given vertex is equal to one:
\begin{eqnarray}
\sum\limits_{i \mbox{-fixed}} n_{ij} = 1.
\label{SumOfOccupationNumbers}
\end{eqnarray}
The idea of effective magnetic field is that the constraint (\ref{SumOfOccupationNumbers}) can be rewritten as a zero divergence of a vector field, which can be referred as magnetic field. If we define the magnetic field along each edge as a linear function of the occupation number (which precise form is determined later), and denote by $B_i(x,y,z)$ its $i$th component in the positive direction from the vertex located at the point with coordinates $x, y, z$, its discretized divergence will be written as:
\begin{eqnarray}
&&B_x(x, y, z) - B_x(x-1, y, z)
\nonumber\\
&&+ B_y(x, y, z) - B_y(x, y-1, z)
\nonumber\\
&&+ B_z(x, y, z) - B_z(x, y, z-1)
=0
\label{DivB}
\end{eqnarray}
Since the last equation contains terms with positive signs at the point with the coordinates $x,y,z$ and terms with negative signs at its neighboring points, whereas the eq. \ref{SumOfOccupationNumbers} contains only terms with positive signs, we have to use the fact that our lattice is bipartite and to introduce the factors $\sigma = \pm 1$ on odd/even sublattice. More specifically, we assume that the magnetic field $B(x,y,z)$ is obtained from the corresponding occupation number through multiplying by $\sigma$. Finally, we need to use the fact that the total sum in the Eq. \ref{DivB} is equal to zero, whereas, in the Eq. \ref{SumOfOccupationNumbers}, the sum is equal to one. In the previous works (e.g. \cite{PhysRevLett.91.167004}), this fact was accounted by subtracting inverse coordination number of the lattice $z$, and, hence the magnetic field was defined as: 
\begin{eqnarray}
B_i (\vec{r}) = \sigma (n_{r , r+e_i} - 1/z ).
\label{OldMagneticField}
\end{eqnarray}

However, the last equation has a drawback: 
the magnetic field does not decay on an infinite trivial lattice, and hence it is hard to use it to compute integral quantities, i.e. the ones, that would become space integrals in the continuum limit, such as e.g. Chern-Simons integral.
 Therefore, we would like to modify the Eq. \ref{OldMagneticField} to resolve these difficulties. 
Specifically, we would like to define $B$ in such a way, that if the dimer configuration is non-trivial only at a finite region of the lattice, the magnetic field would be non-trivial only within that region, and zero away from it. 

\begin{figure}[h]
\centering
\resizebox{8cm}{!}
{
\includegraphics[angle=0]{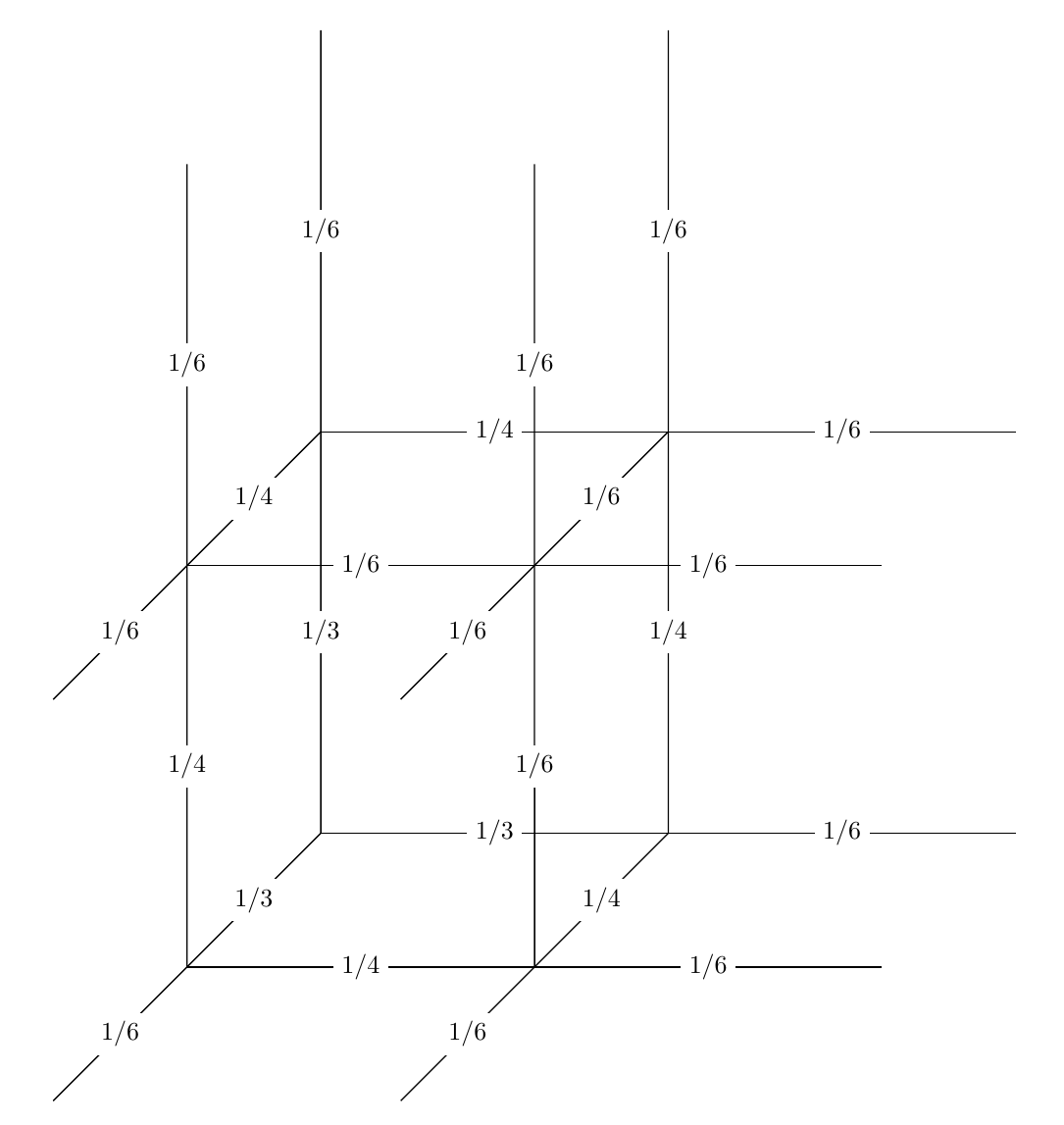}
}
\caption{A possible configuration of  effective coordination numbers $w_{r , r\pm e_i}$ in a lattice dimer model on a cubic lattice. Numerical values of $w_{r , r\pm e_i}$, which satisfy the Eqs. \ref{SumOfWeightsOne}, \ref{LocTransBConstraint},  are shown for each bond. 
}
\label{CoordNumbers}
\end{figure}

We can start from considering a finite lattice, and generalizing the Eq. \ref{OldMagneticField} to be applicable to it. It is easy to see, that on a finite lattice, the Eq. \ref{OldMagneticField} with the coordination numbers taken from the infinite lattice, will not lead to zero divergence at the points located on the boundary points of the lattice. Therefore, we would like to find new quantities instead of the inverse coordination number $1/z$, such that divergence of the new magnetic field is zero everywhere. It is easy to denote them by unknown quantities $w_k$, write the new magnetic field expression as:
\begin{eqnarray}
B_i (\vec{r}) = \sigma (n_{r , r+e_i} - w_{r , r+e_i} ),
\label{NewMagneticField}
\end{eqnarray}
and determine the unknown quantities $w_{r , r+e_i}$. It is possible to fix the values of $w_{r , r+e_i}$  by applying a few conditions. The first condition is that  divergence of the magnetic field is zero at all points including corners and wedges, and it results in the following constraint:
\begin{eqnarray}
\sum\limits_{i} w_{r , r\pm e_i} = 1.
\label{SumOfWeightsOne}
\end{eqnarray}
This equation does not fix the weights $w_{r , r+e_i}$ uniquely. Therefore, without loss of generality, we can impose additional constraints, that 'average' weight over a plaquette is zero:
\begin{eqnarray}
&&  w_{r,\; r+e_x} -w_{r+e_y,\: r+ e_y + e_x} 
\nonumber\\
&& - w_{r+e_z,\:r+ e_z + e_x}  + w_{r+ e_y + e_z,\:r+ e_y + e_z + e_x} = 0
\nonumber\\
&& \mbox{and its cyclic permutations}
\label{LocTransBConstraint}
\end{eqnarray}
We will discuss the precise meaning of these constraints later, but for now, we note that they have a simple interpretation: in a trivially aligned configuration of dimers, magnetic flux over each plaquette is zero. In fact, even in the presence of all these constraints (\ref{SumOfWeightsOne}) and (\ref{LocTransBConstraint}), the weights are still non-unique, but we can select just one arbitrary configuration of  $ w_{r , r\pm e_i}$, satisfying all the equations. A possible example for a finite lattice is presented on the Fig.  \ref{CoordNumbers}. We note, that by fixing $w_{r , r\pm e_i}$ near the corner of the lattice, we can fix them at all other edges of the finite lattice. 

In the case of an infinite lattice, we can view our model in the following way. Suppose that dimers are allowed to form non-trivial configurations only within a finite subregion (we can assume, that it has even number of sites in each direction), whereas at the rest of the lattice, dimers are fixed in a trivial configuration. Suppose also, that within the finite sublattice, the weights are fixed as we discussed previously, e.g. as shown on the Fig. \ref{CoordNumbers}, whereas at the rest of the lattice, the weights are fixed simply as $w_{r , r\pm e_i} = n_{r , r\pm e_i}$ - this is an allowed choice because, we are assuming now, that the dimer configuration does not evolve outside the finite subregion. We also note that our configuration does not break any of the Eqs. (\ref{SumOfWeightsOne}, \ref{LocTransBConstraint}), because dimers reside either within the finite sublattice or away from it, and therefore at the interface we have  $w_{r , r\pm e_i} = n_{r , r\pm e_i} = 0$ - this does not affect the Eqs. (\ref{SumOfWeightsOne}, \ref{LocTransBConstraint}). 
In this configuration, the lattice magnetic field is non-trivial only within the finite sublattice, and zero outside of it, particularly at infinity.


\section{Lattice Hopf number}
\label{EffectiveVectorPotentialAndLatticeHopfNumber}

\begin{figure}[h]
\centering
\resizebox{4cm}{!}
{
\includegraphics[angle=0]{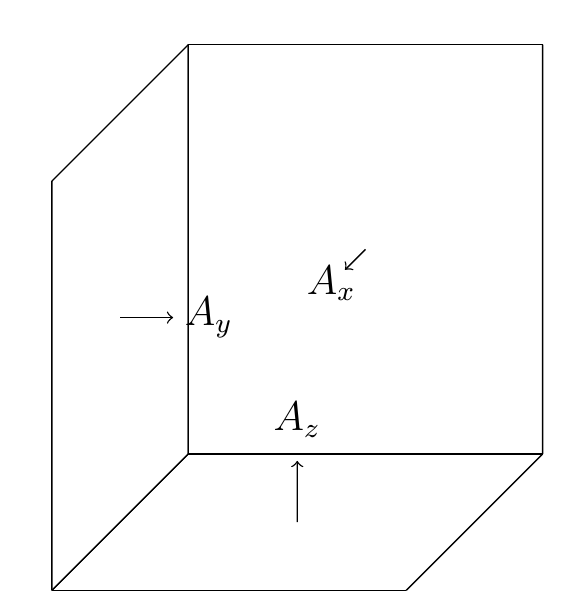}
}
\caption{Each component of vector potential in lattice dimer model is defined perpendicularly to the corresponding plaquette.}
\label{VectorPotentialFig}
\end{figure}

Once we have defined the lattice magnetic field (Eq. \ref{NewMagneticField}) and fixed the weights, we can also define lattice vector potential. Since the magnetic field is defined along each edge, the corresponding vector potential will be defined at each plaquette (see Fig. \ref{VectorPotentialFig}), through the equations:
\begin{eqnarray}
B_x (x,y,z) &=& A_z (x,y,z) - A_z(x,y-1,z)
\nonumber\\
&& - A_y(x,y,z) + A_y (x,y,z-1),
\label{VectorPotential}\\
B_{y,z} &&\mbox{are defined through cyclic permutations}
\nonumber
\end{eqnarray}

Using these equations together with gauge-fixing conditions, we can find the vector potential for a given configuration of the magnetic field. For example, if we fix the gauge $A_z = 0$, we can write the vector potential as:
\begin{eqnarray}
&& A_x (x,y,z) = \sum\limits_{k=1}^{z} B_y (x,y,k),
\nonumber\\
&& A_y (x,y,z) = -\sum\limits_{k=1}^{z} B_x (x,y,k).
\label{VectorPotentialExpression}
\end{eqnarray}

From the Eqs. (\ref{VectorPotential}), it is easy to see that a local dimer flip, shown on the Fig. \ref{DimerFlip} can be written as a change of vector potential at the plaquette, where the flip occurs. Indeed, the change of vector potential at one plaquette results in change of magnetic field only within the bonds surrounding such plaquette, and it is straightforward to check that if the bonds contain two parallel dimers, then unit change of the vector potential corresponds to unit change of effective magnetic fields around the plaquette, which is equivalent to a dimer flip.

Once we have defined the vector potential, we can use it to define Hopf number. It is known, that in a continuum theory, Hopf invariant can be written as a Chern-Simons integral, i.e. an integral of a vector field $A$ and its rotor $B$  \cite{PhysRevLett.51.2250, Volovik1977}:
\begin{eqnarray}
\chi = \int d^3 x \vec{A} \vec{B},
\label{ContinuumHopfNumber}
\end{eqnarray}
We would like to demonstrate, that, in the lattice dimer model, we can define a similar expression in terms of the effective vector potential and magnetic field, which is equal to an integer number, and remains invariant under any local plaquette flip. 

First, we have to discretize the Eq. (\ref{ContinuumHopfNumber}) properly. We do it by replacing the integral with a sum over the values of vector potential over each plaquette, multiplied by magnetic field averaged over all edges, attached to the plaquette:
\begin{eqnarray}
\chi &=& \sum\limits_{x,y,z} 
\frac{A_x (x,y,z) }{8} 
\label{DiscreteHopfNumber}
\\
& \times &
\left( 
B_x(x,y,z) + B_x (x,y+1,z) 
\right. \nonumber\\ && \left.
+ B_x (x,y,z+1) + B_x (x,y+1,z+1)
\right. \nonumber\\ && \left.
+ B_x(x-1,y,z) + B_x (x-1,y+1,z) 
\right. \nonumber\\ && \left.
+ B_x (x-1,y,z+1) + B_x (x-1,y+1,z+1)
\right) 
\nonumber\\
&+&( \mbox{cyclic permutations}).
\nonumber
\end{eqnarray}

Next, we would like to check if the last equation satisfies the same properties, as the continuum Hopf number, defined by the Eq. (\ref{ContinuumHopfNumber}). For example, the expression (\ref{ContinuumHopfNumber}) is known to be gauge invariant. Indeed, it is easy to see that, in the case of an infinite space with decreasing field at infinity, its variation is equal to the integral of divergence of the magnetic field, i.e. zero. We demonstrate in the Sec. \ref{Sec:GaugeInvariance}, that the same statement holds for the discrete Hopf number, defined by the Eq. (\ref{DiscreteHopfNumber}). We note, that to prove it, we use the fact that magnetic field is decreasing at the infinity. Furthermore, we note, that the same statement holds for a finite lattice, provided that the magnetic field satisfies the Eq. (\ref{DivB}) at all points.

Now, let us consider the transformation of Hopf number (\ref{ContinuumHopfNumber}) under infinitesimal change of vector potential. Indeed, in continuum field theory, Hopf number transforms as:
\begin{eqnarray}
\frac{\delta \chi}{\delta A_i} = 2 B_i.
\label{deltaChi}
\end{eqnarray}
In the case of the lattice model, we can obtain a similar expression (see \ref{Sec:LocTransA} 
for the derivation):
\begin{eqnarray}
&& \Delta \chi (\Delta A_x (x,y,z)) = 2 A_i (x,y,z) 
 \nonumber\\ 
&& \quad \times  \frac{1}{8}
\left\{  B_x (x,y,z) + B_x (x,y+1,z) 
\phantom{\frac{1}{1}}
\right. \nonumber\\ && \qquad \left.
+ B_x (x,y,z+1) + B_x (x,y+1,z+1)
\right. \nonumber\\ && \qquad \:\:\: \left.
  B_x (x-1,y,z) + B_x (x-1,y+1,z) 
\right. \nonumber\\ && \qquad \left.
+ B_x (x-1,y,z+1) + B_x (x-1,y+1,z+1)
\phantom{\frac{1}{4}} \! \!\! \right\}
\nonumber\\
&&\qquad \mbox{Cyclic permutations for $A_{y,z}$}
\nonumber\\
\label{ChiTransLocA}
\end{eqnarray}
This equation tells us that  if we change the vector potential at one plaquette, the corresponding transformation of the Hopf number $\Delta \chi$ will be expressed as variation of the vector potential, multiplied by average magnetic field at the edges, emerging perpendicularly to the plaquette. If we use the fact that a local dimer flip, shown on the Fig. \ref{DimerFlip}, corresponds to a local change of vector potential, we can conclude that variation of $\chi$ under a plaquette flip is proportional to average magnetic field perpendicular to the plaquette. Since in this configuration, dimer occupation numbers are non-zero only along the plaquette, we can conclude that $\Delta \chi$ is proportional to averaged weights of edges emerging perpendicularly to the plaquette, i.e. precisely the combinations entering the Eq. (\ref{LocTransBConstraint}). Thus, we arrive to the conclusion: the constraints (\ref{LocTransBConstraint}) result in the Hopf number $\chi$ being invariant under any local dimer flips. 

Once we have established that Hopf number (\ref{DiscreteHopfNumber}) is an invariant, we are interested in computing it explicitly. The simplest field configuration, where it can be computed, is a trivial dimer configuration, where all dimers are aligned in one direction, forming maximally flippable state. In this case, after fixing the gauge $A_z = 0$, straightforward applying of the constraint (\ref{LocTransBConstraint}) to each plaquette, leads to the conclusion that $\chi = 0$.

\begin{figure}[h]
\centering
\resizebox{8cm}{!}
{
\includegraphics[angle=0]{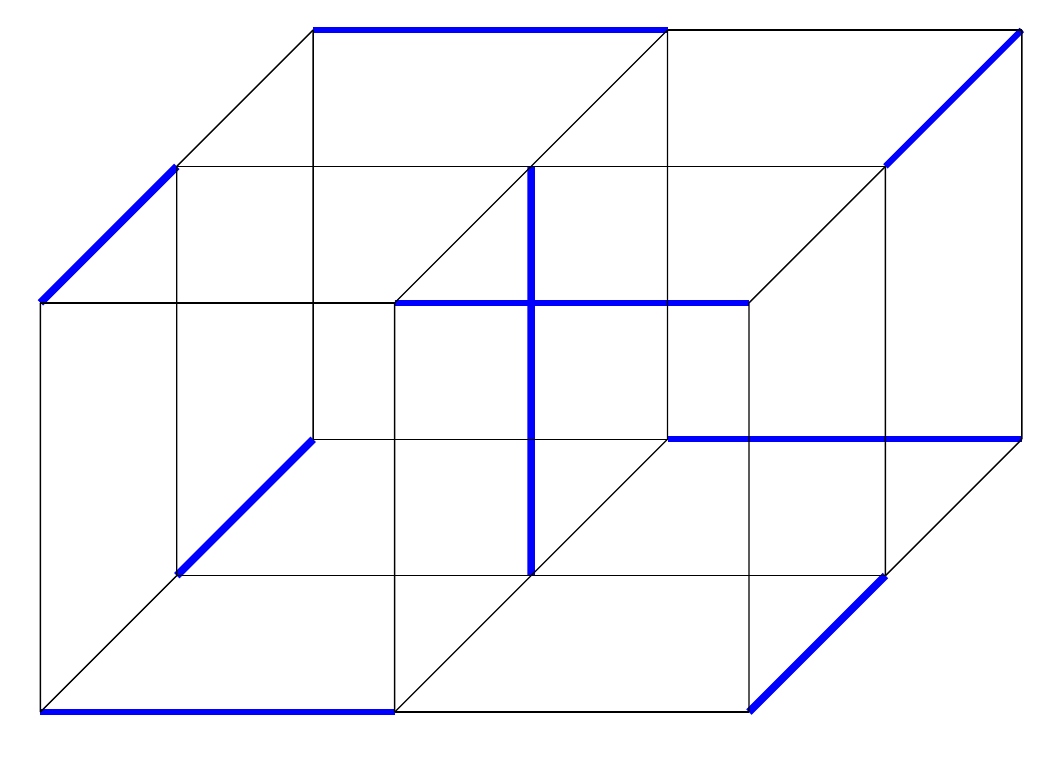}
}
\caption{Dimer configuration forming a hopfion - the simplest topologically non-trivial configuration }
\label{JustHopfion}
\end{figure}

From the Ref. \cite{PhysRevB.84.245119}, we know the simplest topologically non-trivial configuration of dimers (hopfion), which we show on the Fig. \ref{JustHopfion}. In \cite{PhysRevB.84.245119} it was shown, that such configuration is topologically non-trivial, because pfaffian of its Kasteleyn matrix is equal to $-1$, in contrast to $+1$ for the trivial, maximally flippable state, and it does not change under local flips. It is easy to demonstrate (see \ref{Sec:ExplicitCalculationOfHopfNumber} for details), that such non-trivial configuration has, indeed, $\chi = 1$. It is also easy to see, that the Hopf number remains invariant if the hopfion is placed on a larger lattice with trivially aligned dimers. Furthermore, if  several hopfions are placed on a large lattice filled with trivially aligned dimers, their total Hopf number is given by the sum of Hopf numbers for each hopfion. These facts confirm an idea, that the configuration, shown on the Fig. \ref{JustHopfion}, has all properties of conventional topological defects. At the same time, it is distinguished by the fact, that its topological properties are exact on the lattice, i.e. do not require taking continuum limit. 

In conclusion, we have demonstrated that 3D cubic lattice dimer model possesses topological defects characterized by an integer Hopf number. The precise meaning of the word 'topological' is that the Hopf number remains invariant under local flips, as we showed on the Fig.  \ref{DimerFlip}. 


\section{Discussion}
\label{Discussion}

In this paper, we have demonstrated that 
configurations in 3D bipartite lattice dimer model can be classified according to their Hopf numbers, which are expressed in terms of effective magnetic field of the dimer model. These Hopf numbers are preserved under local dimer flips (Fig. \ref{DimerFlip}), and therefore configurations with different Hopf numbers cannot be transformed into each other. 


Throughout the paper we were assuming that the lattice is cubic, but we would like to emphasize that our reasoning works for any 3D bipartite lattice. Indeed the only necessary property of the lattice we used, is the existence of effective magnetic field, which can be defined on any bipartite lattice.

 Furthermore, we were assuming that the dimers are constrained by a condition (Eq. \ref{SumOfOccupationNumbers}) -  exactly one dimer is attached to each vertex. However, in a similar way, Hopf number can be defined for a lattice dimer model, in which the number of dimers attached to each vertex is fixed to be two. Indeed, the 'conventional' definition of effective magnetic field (similarly to the eq. \ref{OldMagneticField}) will become:
\begin{eqnarray}
B_i (\vec{r}) = \sigma (n_{r , r+e_i} - 2/z )
\nonumber
\end{eqnarray}
and to find the weights, we will have to replace the Eq. (\ref{SumOfWeightsOne}) with a new constraint:
\begin{eqnarray}
\sum\limits_{i} w_{r , r\pm e_i} = 2.
\nonumber
\end{eqnarray}
The rest of the derivation will remain the same. 

Our results may have a wide range of applications. It is known, that dimer models on bipartite lattices are dual to spin systems, and, particularly, dimer model on a diamond lattice is dual to spins on pyrochlore lattice, i.e. spin ice \cite{PhysRevLett.96.097207,  PhysRevB.69.064404, 0034-4885-77-5-056501}. Thus, it might be of interest to search for hopfions in various spin systems. Furthermore, since classical spin ice has been proven to exist in such materials as $\mathrm{Dy_2 Ti_2 O_7}$ and $\mathrm{Ho_2 Ti_2 O_7}$ \cite{RevModPhys.82.53, 0034-4885-77-5-056501}, it would be interesting to search for hopfions experimentally. We note that in the realistic spin ice mateials, one can consider two possible regimes. Without external magnetic field, the spins form 2-in, 2-out magnetic order, which is dual to a lattice dimer model with two dimers 'touching' each vertex. However, in the presence of external magnetic field in $[111]$ direction, the spins may form 3-in, 1-out magnetic order \cite{BrianYee2016}, which is dual to the lattice dimer model with one dimer 'touching' each vertex considered here. In the future, we are interested in studying this question more deeply, e.g. computing relevant physical observables affected by hopfions. 

Experimental discovery of hopfions may
lead to plenty of novel phenomena. Indeed, topological defects described by a Hopf number has been studied in various physical contexts, and cores of hopfions were predicted, for instance, to host non-abelian anyons \cite{PhysRevB.84.184501, PhysRevB.84.245119}, which are interesting in the context of quantum computing \cite{pachos_2012}. For this reason, it is of interest to explore possible implications of hopfions in our context. For instance, it is of interest to explore if hopfions in spin ice can host non-abelian anyons or other kinds of localized states. 

During the recent years, there have been ongoing attempts to study magnetic frustration in 
artificial spin ice \cite{RevModPhys.85.1473, Perrin2016, Lao2018}, where the dimers are simulated by nanomagnets. Indeed, in \cite{Perrin2016}, a successful creation of frustrated artificial spin ice on quadratic lattice was reported. Since, in this model, the nanomagnets obey the same ice rule as in classical spin ice, it seems plausible to use artificial spin ice for realizing hopfions. Strictly speaking, in this case, there appears an issue, such that dynamics of nanomagnets does not necessarily gets reduced to local dimer flips. Instead, configurations in ASI evolve through creation/annihilation of monopole-antimonopole pairs. However, this problem does not make 
realization of hopfions impossible: ASI dynamics would get reduced to local dimer flips if monopoles are allowed to travel only within one plaquette, which may happen if the monopole-antimonopole pairs quickly annihilate after creation. Furthermore in \cite{Farhan2013}, simultaneous switching of magnetization directions around a hexagon in kagome spin ice was reported, which is exactly equivalent to local dimer flips around a hexagonal plaquette. We believe that further steps in this direction, e.g. realizing ASI on a cubic lattice with simultaneous magnetization flips around each square plaquette, may result in realizing ASI, which dynamics is the same as considered in this paper. Finally, we mention, that in the recent years, there were numerous efforts to create 3D artificial spin ice \cite{doi:10.1063/1.4861118, FernGUndez-Pacheco2017, Keller2018}, and we emphasize that creating hopfions requires a lattice with just two layers. 

In conclusion, we believe that in should be possible to realize hopfions in artificial spin ice, once it will become possible to realize dynamics through local flips on a 3D lattice.

\begin{acknowledgments}   
The author would like to thank professor R. Melko for multiple discussions on the project and providing valuable feedback. The author also would like to thank Dr. F. Levkovich-Maslyuk, J. Rau, S. Syzranov, C. Nisoli, R. Moessner for helpful discussions on the project. This work was supported by NSERC of Canada. 
\end{acknowledgments}   


\appendix
\section{}

In this appendix we present derivation of the properties of the discrete Hopf number, defined according to the Eq. (\ref{DiscreteHopfNumber}) of the main text. For shortness of notations, we find it convenient to represent the sum of products of magnetic fields and vector potentials (\ref{DiscreteHopfNumber}) graphically. Particularly, we assume that summation is taken over all plaquettes, and denote vector potential in each of them by red square. We also denote magnetic fields, which are multiplied by the vector potential,  by green lines. If we view a red square with attached green lines as a corresponding vector potential multiplied by the adjacent magnetic field, we can represent our Hopf number, defined by the Eq. (\ref{DiscreteHopfNumber}), graphically as: 

\vspace{-0.5cm}

\begin{eqnarray}
	\includegraphics[scale=0.8, angle=0] 
	{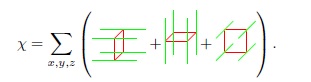}
\nonumber
\end{eqnarray}

In the following sections, we use these graphical notations to derive its gauge invariance, and transformation under smooth variations of the vector potential, i.e. the eq. (\ref{ChiTransLocA}) of the main text.


\subsection{Gauge invariance}
\label{Sec:GaugeInvariance}

In this section, we demonstrate, that the discrete Hopf number, defined by Eq. (\ref{DiscreteHopfNumber}) of the main text is invariant under gauge transformations of the vector potential, i.e. transformations which leave the magnetic field invariant. Indeed, similarly to conventional electrodynamics, the lattice magnetic field remains invariant, if the vector potential is transformed as:
\begin{eqnarray}
A_i (\vec{r}) \to A_i (\vec{r}) + \theta(\vec{r}) - \theta(\vec{r}-\vec{e_i}).
\label{AGaugeTrans}
\end{eqnarray}
On the lattice, the gauge function $\theta$ can be defined in every cube formed by the lattice sites, is such a way that the vector potential on a given plaquette is added by a difference between the gauge functions at cubes adjacent to the plaquette. In other words, if we denote the gauge function at each lattice unit by a yellow cube, we can represent the gauge transformation (\ref{AGaugeTrans}) graphically as:

\vspace{-0.5cm}
\begin{eqnarray}
	\includegraphics[scale = 0.8, angle=0] 
	{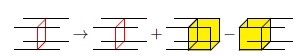}
\nonumber
\end{eqnarray}
\vspace{-0.5cm}

In these notations, variation of the discrete Hopf number under gauge transformations of the vector potential (which is a sum of products between gauge functions and magnetic fields) can be represented graphically as:

\vspace{-0.5cm}
\begin{eqnarray}
	\includegraphics[scale=0.8, angle=0] 
	{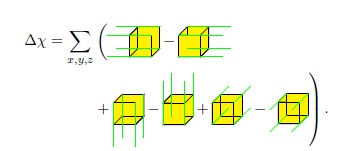}
\label{ChiTransform}
\end{eqnarray}
\vspace{-0.5cm}

Now we have to use the fact that our lattice magnetic fields are non zero only within the finite region of the total lattice, which implies that the terms entering the Eq. (\ref{ChiTransform}) are non zero only within the finite sublattice. The fact that the terms entering the sum (\ref{ChiTransform}) are zero away from the finite sublattice, implies that we can shift the sum, in such a way that in each bracket of the Eq. (\ref{ChiTransform}), the gauge function is taken at one point. Therefore, we can rewrite the gauge transformation as:

\vspace{-0.5cm}
\begin{eqnarray}
	\includegraphics[scale = 0.8, angle=0] 
	{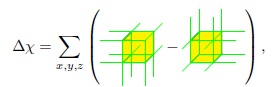}
\nonumber
\end{eqnarray}
\vspace{-0.5cm}

which is just a sum of divergences of magnetic fields at all corners of the cube entering each bracket. Since the divergence of magnetic field is zero, we can conclude that $\Delta \chi$ is zero, and thus the discrete Hopf number $\chi$ is gauge invariant.


\subsection{Local transformations of vector potential}
\label{Sec:LocTransA}

In this section we derive the transformation of the discrete Hopf number under local change of the vector potential, i.e. change of  $A_z(x,y,z)$ at one plaquette. The latter results in change of $B_{y, z}$ at the edges, adjacent to the plaquette. If we denote by dashed lines the bonds, where the vector potential or magnetic field is varied, we can represent the terms contributing to the change of Hopf number as:

\vspace{-0.5cm}
\begin{eqnarray}
	\includegraphics[scale=0.8, angle=0] 
	{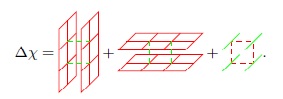}
\nonumber
\end{eqnarray}

Here, the first two graphs show the plaquettes, whose vector potential we have to include, and variations of magnetic fields, by which they have to be multiplied. The last graph shows the plaquette with varying vector potential multiplied by the adjacent magnetic fields. 

In the last equation we can express the variations of magnetic field in terms of variations of the vector potential. If we do it with the first two terms, we can rewrite it as:

\vspace{-0.5cm}
\begin{eqnarray}
	\includegraphics[scale=0.8 ,angle=0] 
	{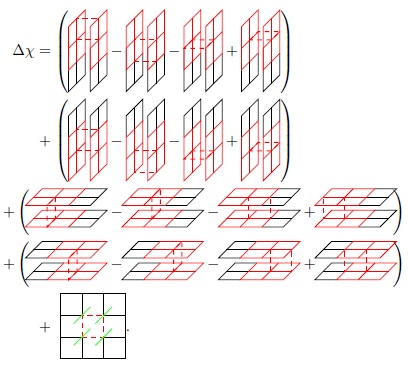}
\nonumber
\end{eqnarray}

If we similarly express the magnetic field in the last term, the variation of the discrete Hopf number will be simplified as:

\vspace{-0.5cm}
\begin{eqnarray}
	\includegraphics[scale=0.8 , angle=0] 
	{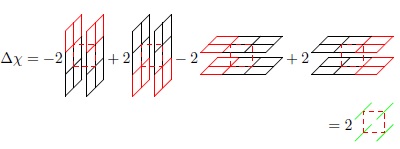}
\nonumber
\end{eqnarray}

This equation is rewritten without graphic notations precisely as the Eq. (\ref{ChiTransLocA}) from the main text. 


\subsection{Explicit calculation of Hopf number}
\label{Sec:ExplicitCalculationOfHopfNumber}

\begin{figure}[h]
	\includegraphics[scale=0.9 , angle=0] 
	{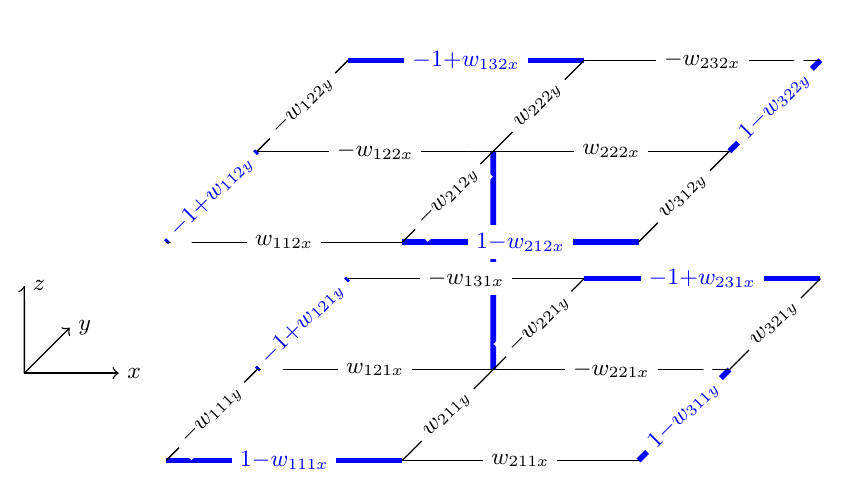}
\vspace{-0.25cm}
\caption{\small Values of the effective magnetic field along the bonds forming the hopfion. }
\label{HopfionMagneticFields}
\end{figure}

In this section, we present explicit calculation of the Hopf number (Eq. \ref{DiscreteHopfNumber} from the main text) for the hopfion configuration, shown on the Fig. \ref{JustHopfion}. We assume that the hopfion is placed on a lattice, where all other dimers are aligned in $z$ direction. Without loss of generality, we can fix the gauge as:  $A_z = 0$. Since the weights satisfy the condition (\ref{LocTransBConstraint}), it is easy to see that aligned dimers result in zero average magnetic field over plaquette, and thus the Hopf number is contributed only by plaquettes and edges 'touching' the hopfion. Explicitly,  relevant edges have values of magnetic field shown on the figure \ref{HopfionMagneticFields}.

In this section we choose indices in the following way: three numbers refer to coordinates of the lattice site, from which the edge starts in the positive direction, and $xyz$ refer to the edge's direction.

\begin{figure}[h]
	\includegraphics[scale=0.9 , angle=0] 
	{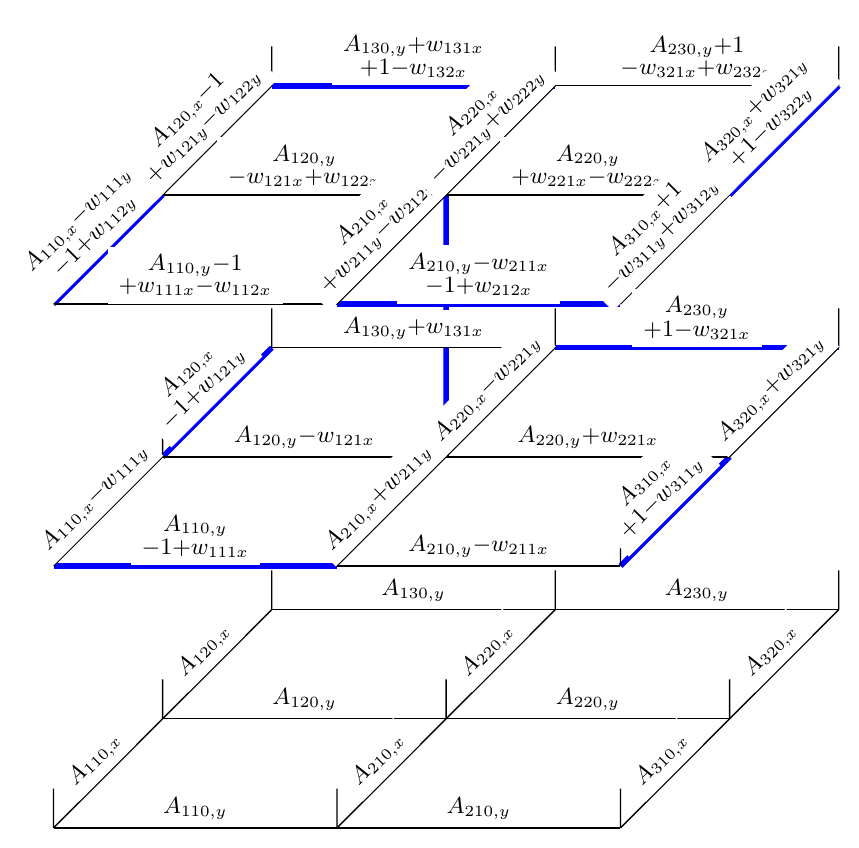}
\vspace{-0.25cm}
\caption{\small Values of the effective vector potential on the plaquettes forming the hopfion. Since we fixed the gauge $A_z = 0$, we show only vector potentials on vertical plaquettes (i.e. only $A_{x,y}$).}
\label{HopfionVectorPotential}
\end{figure}

After fixing the gauge $A_z$, we can express the values of vector potential at arbitrary $z$ in terms of the values of vector potential 'below' the hopfion, i.e. at $z=0$. The resulting values of vector potential at the plaquettes, contributing to the Hopf invariant, are shown on the Fig. \ref{HopfionVectorPotential}.

\vspace{-0.5cm}

\begin{figure}[h]
	\includegraphics[scale=0.9 , angle=0] 
	{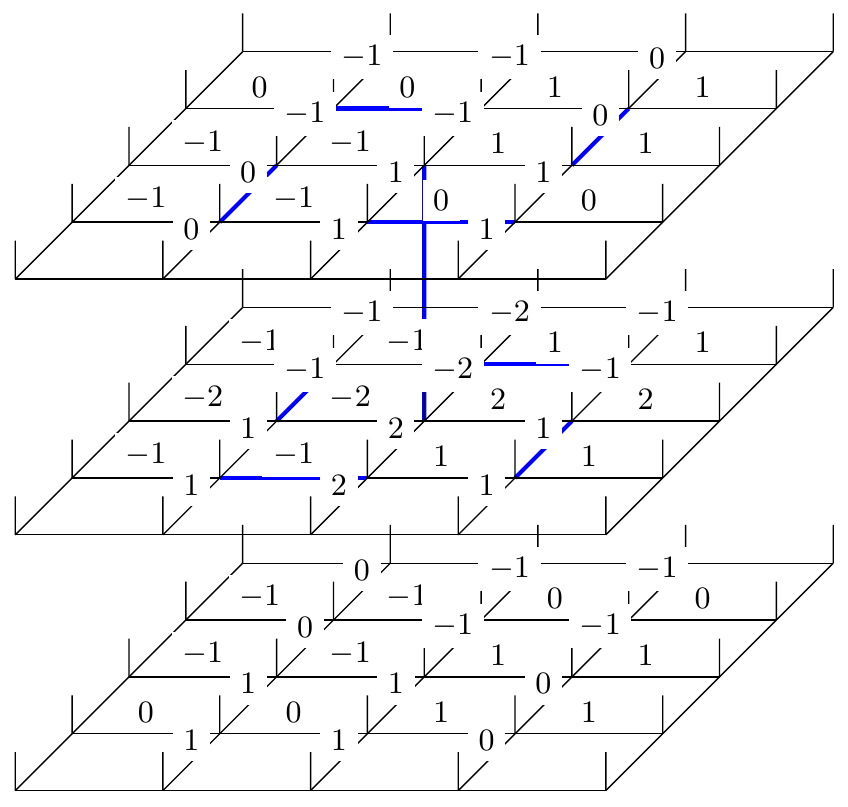}
\caption{\small Values of the sum of magnetic fields over bonds adjacent to each plaquette. In other words, a number at each plaquette represents the sum  $\sum B_{x,y}$ over each plaquette, as it enters the Eq. (\ref{DiscreteHopfNumber}) from the main text.
These numbers have to be multiplied by the components of vector potential at the same plaquettes and summarized in order to obtain the Hopf invariant. 
}  
\label{HopfionSumOfMagneticFields}
\end{figure}

Once, we have the values of $\vec{A}$ and $\vec{B}$, application of the Eq. (\ref{DiscreteHopfNumber}) is straightforward. It is simplified by the fact, that when we compute a sum of magnetic fields 'piercing' a plaquette, the coordination numbers cancel out, and, as a result, such sum is equal to a number of  'horizontal' dimers in a hopfion taken with appropriate sign. Explicitly, the sum of magnetic fields 'piercing' each plaquette is shown on the Fig. \ref{HopfionSumOfMagneticFields}.

 We assume here, that the hopfion is placed in a space with trivially aligned dimers, but the result would not change if two hopfions 'touch' each other. One can check that, in the latter case, the contribution from plaquettes lying between the hopfions can be split between them, so that each hopfion acruires the same contribution, as if it were in a vacuum.   

\begin{figure}[h]
\centering
\begin{subfigure}[h]{0.5\textwidth}
\includegraphics[scale=1, angle=0] 
	{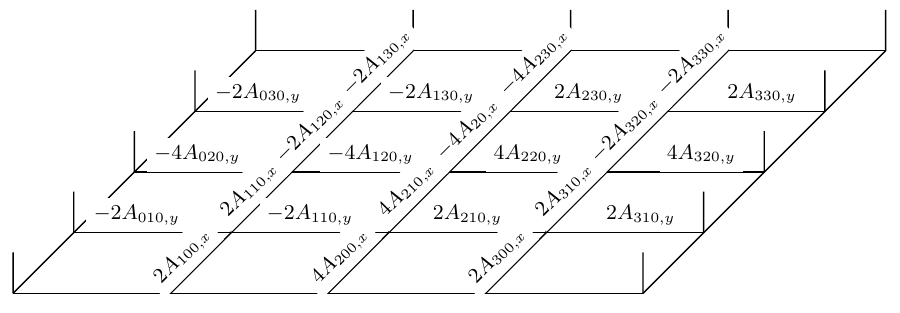}
\vspace{-0.2cm}
\caption{}
\label{HopfNumTerms1}
\vspace{-0.2cm}
\end{subfigure}
\begin{subfigure}[h]{0.5\textwidth}
\includegraphics[scale=1, angle=0] 
	{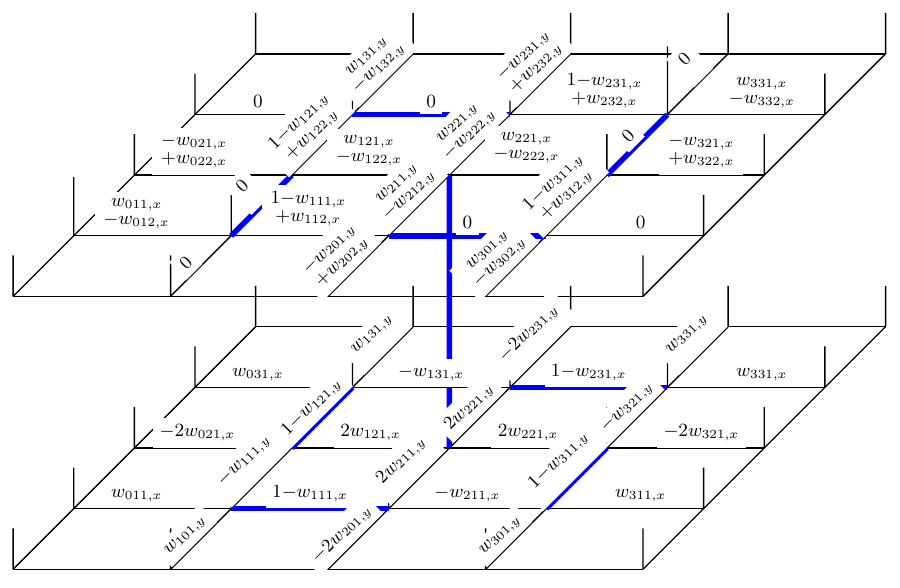}
\vspace{-0.2cm}
\caption{}
\label{HopfNumTerms2}
\vspace{-0.2cm}
\end{subfigure}
\caption{\small Hopf invariant can be expressed as one eighth of the total sum of the terms shown on the Figs. \ref{HopfNumTerms1}
, \ref{HopfNumTerms2}: $\chi = \frac{1}{8} \times \sum \left( \mathrm{terms \; from \; Fig.\: \ref{HopfNumTerms1}} \right) +  \frac{1}{8} \times \sum \left( \mathrm{terms \;  from \;  Fig.\: \ref{HopfNumTerms2}} \right)$. We presented these terms graphically in order to demonstrate the most convenient way of summaraizing them.
 }
\label{AllTermsHopfInvariant}
\end{figure}

After we multiply each component of vector potential by the corresponding sum of magnetic fields, we can see that all terms contributing to the Hopf invariant can be split into two groups: terms arising from the vector potential 'below' the hopfion, i.e. proportional to $A_{ij0, x,y}$, and terms arising from coordination numbers at its edges. We represent all of these terms graphically on the Fig. \ref{AllTermsHopfInvariant}. More specifically, the Fig. \ref{HopfNumTerms1} shows all terms arising from the vector potential $A_{ij0, x,y}$, which can be obtained from the corresponding contributions to the total vector potential (shown on the Fig. \ref{HopfionVectorPotential}) by multiplying over the sum of magnetic fields (see Fig. \ref{HopfionSumOfMagneticFields}) and summing over the planes. In contrast, the Fig. \ref{HopfNumTerms2} shows all terms arising from the coordination numbers - the corresponding contributions to the total vector potential from the Fig.  \ref{HopfionVectorPotential} also multiplied by the sum of magnetic fields. Thus, the total Hopf invariant is equal to one eighth of the total sum of terms shown on the Figs. \ref{HopfNumTerms1}, \ref{HopfNumTerms1}.




\begin{figure}[h]
\centering
\begin{subfigure}[h]{0.5\textwidth}
\includegraphics[scale=1, angle=0] 
	{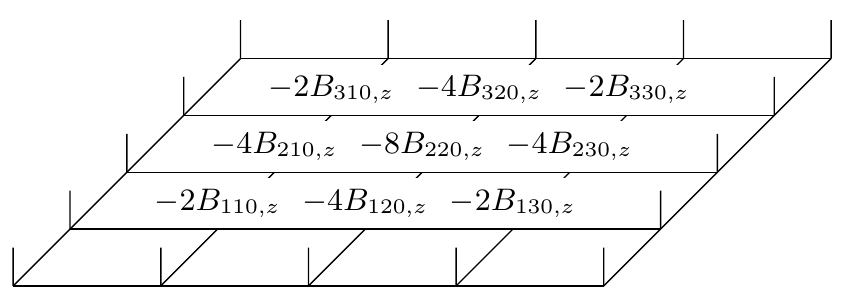}
\vspace{-0.2cm}
\caption{}
\label{HopfNumEq2Terms1}
\vspace{-0.2cm}
\end{subfigure}
\begin{subfigure}[h]{0.5\textwidth}
\includegraphics[scale=1, angle=0] 
	{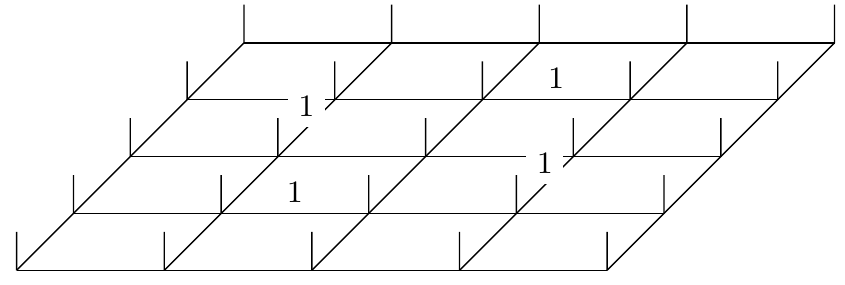}
\vspace{-0.2cm}
\caption{}
\label{HopfNumEq2Terms2}
\vspace{-0.2cm}
\end{subfigure}
\begin{subfigure}[h]{0.5\textwidth}
\includegraphics[scale=1, angle=0] 
	{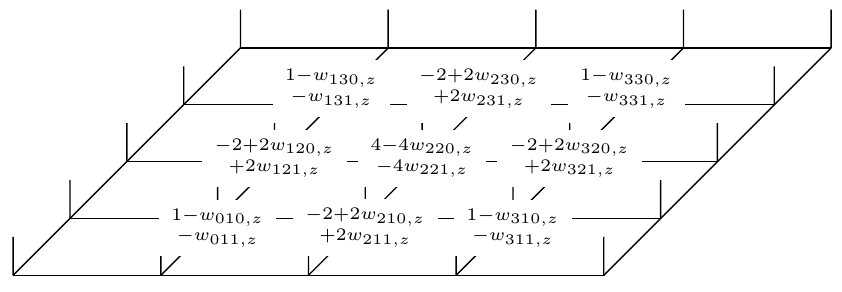}
\vspace{-0.2cm}
\caption{}
\label{HopfNumEq2Terms3}
\vspace{-0.2cm}
\end{subfigure}
\caption{\small Hopf invariant is equal to a superposition of the terms shown here: 
\newline
$\chi = \frac{1}{8} \times \left( \mbox{Sum \; of \; terms \; from \; Fig.\: \ref{HopfNumEq2Terms1} } \right) + \frac{1}{4} \times \left( \mbox{Sum \; of \; terms \; from \; Fig.\: \ref{HopfNumEq2Terms2} } \right)
+ \frac{1}{8} \times \left( \mbox{Sum \; of \; terms \; from \; Fig.\: \ref{HopfNumEq2Terms3} } \right) $.
\newline
 In particular, the Fig. \ref{HopfNumEq2Terms1} shows all terms obtained by summing the terms from Fig. \ref{HopfNumTerms1}. Fig. \ref{HopfNumEq2Terms2} shows all terms obtained by summing the terms from the top layer of the Fig. \ref{HopfNumTerms2}. Fig. \ref{HopfNumEq2Terms3} shows all terms obtained by summing the terms from the bottom layer of the Fig.  \ref{HopfNumTerms2} 
 }
\end{figure}

From the Fig. \ref{HopfNumTerms1} one can see, that the components of $A_{x,y}$ sum up in such a way, that they give magnetic fields 'below' the hopfion, as we show on the Fig. \ref{HopfNumEq2Terms1}. 
The contributions from weights in the top layer of the hopfion (see the top layer of the Fig. \ref{HopfNumTerms2} ) cancel out due to the Eq. (\ref{LocTransBConstraint}). Therefore, the total contribution from the top layer of the Fig. \ref{HopfNumTerms2} gives just four ones coming from different vortices, as we show on the Fig. \ref{HopfNumEq2Terms2}. On the other hand, the weights from the bottom layer of the Fig. \ref{HopfNumTerms2} can be transformed by applying the condition of zero divergence (\ref{SumOfWeightsOne}), thus resulting in the terms shown on the Fig. \ref{HopfNumEq2Terms3}. 
  The total Hopf number can be rewritten as a sum:
\begin{eqnarray}
\chi = \frac{1}{8} \times \left( \mbox{Sum \; of \; terms \; from \; Fig.\: \ref{HopfNumEq2Terms1} } \right) 
\nonumber\\
+ \frac{1}{4} \times \left( \mbox{Sum \; of \; terms \; from \; Fig.\: \ref{HopfNumEq2Terms2} } \right)
\nonumber\\
+ \frac{1}{8} \times \left( \mbox{Sum \; of \; terms \; from \; Fig.\: \ref{HopfNumEq2Terms3} } \right)
\nonumber
\end{eqnarray}

One can check that, since the bonds 'below' the hopfion do not contain dimers, the magnetic field components shown on the Fig. \ref{HopfNumEq2Terms1} cancel out due to the constraint (Eq. \ref{LocTransBConstraint}). Similarly, the weights shown on the Fig. \ref {HopfNumEq2Terms3} cancel as well. Thus, we arrive to the final answer for the Hopf invariant:
\begin{eqnarray}
\chi = 1.
\end{eqnarray}



\bibliographystyle{apsrev4-1}
\bibliography{HopfLinksBib}

\end{document}